# Reaching 5E-13 short term frequency stability of the integrating sphere cold atom clock

**P Liu, Y L Meng, J Y Wan, X M Wang, Y N Wang, L Xiao, H D Cheng\* and L Liu†**

Key Laboratory of Quantum Optics and Center of Cold Atom Physics,
Shanghai Institute of Optics and Fine Mechanics, Chinese Academy of Sciences,
Shanghai 201800, China
E-mail: †liang.liu@siom.ac.cn, \*chenghd@siom.ac.cn

**Abstract.** We present an improvement of short term frequency stability of the integrating sphere cold atom clock after increasing the intensities of clock signals and optimizing the feedback loop of the clock. A short term frequency stability of $5.0 \times 10^{-13} \tau^{-1/2}$ has been achieved and the limiting factors have been analyzed.

## 1. INTRODUCTION

The integrating sphere cold atom clock (ISCAC) has potential applications in space because of its compact volume and great performance in microgravity environment [1]. For an ISCAC, atoms are cooled by diffuse laser light [2,3], which makes it simple and robust because any careful alignment of laser beam and light polarization adjustment are not necessary. Here we use a cylindrical microwave cavity whose inner surface was sanded and electroplated with a layer of silver to meet the requirement of diffuse reflectance. The cooling lights are directly injected into the microwave cavity and the diffuse light is formed by the reflection of the inner sanded surface. The microwave cavity serves two functions for the cold atom clock, that is, cooling atoms and microwave interrogation with cold atoms. The frequency stability of $7.3 \times 10^{-13} \tau^{-1/2}$ has been achieved with this setup and the limiting factors have been analyzed in our previous work [4]. It is found that the main noise sources are from the AM and PM noises of the probe laser, the feedback electronics noise, the atomic shot noise, and the local oscillator phase noise. In order to improve the frequency stability further, we reduce the effect of AM and PM noises by improving the intensities of clock signals. Furthermore, the feedback electronics noise is reduced by changing the feedback loop of the ISCAC.

Previously, the error signals for frequency correction were directly sent to the local oscillator (LO) by an analog circuit. The electronic noises of the feedback loop degraded the frequency stability of the ISCAC. In this work, the error signals are sent to the frequency auxiliary output generator which is phase locked to the LO.

In this paper, the first part is dedicated to generally describe the clock system. Then, we present the improvement recently made on the probe laser and the servo loop for the LO. The third section is devoted to show the improved performance of the ISCAC.

## 2. THE SETUP AND THE IMPROVMENT ON THE CLOCK

The core of the ISCAC setup is the cylindrical microwave cavity which is also used for cooling atoms. Figure 1 shows the schematic of the ISCAC's physical package. The microwave cavity has a loaded quality factor of about 11000 and is adjusted to resonate at the transition frequency of 6.834 GHz between the two rubidium atom's ground states on the $TE_{011}$ electromagnetic mode. The cavity is placed in a vacuum chamber which is covered by an Al barrel. The solenoid coils intertwining around the barrel is used to provide the bias magnetic field. All these devices are covered by five-layer of magnetic shields to attenuate the external magnetic fields.

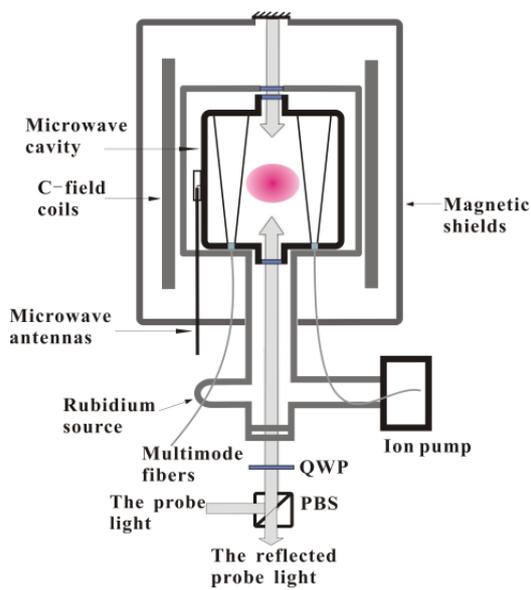

**Figure 1.** Schematic of the physical package. QWP-quarter wave plate, PBS-polarizing beam splitter.

The cooling, pumping, and repumping lights are injected into the microwave cavity with four multimode fibers set in the bottom of the cavity [5]. The probe light propagates along the axis of the cylindrical microwave cavity and is retroreflected by a mirror horizontal set on the top of the physical package. The standing wave probe light is used to avoid heating up the cold atoms.

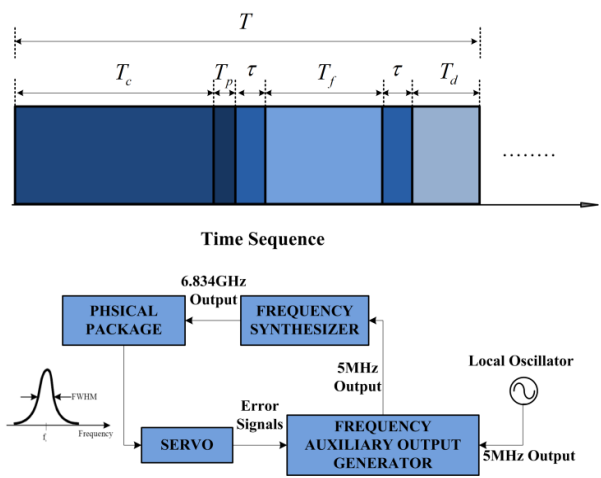

**Figure 2.** The time sequence of operating the integrating sphere cold atom clock and the architecture of the frequency lock loop. Where T=133 ms, $T_c$=95 ms, $T_p$=0.5 ms, $\tau$=3 ms, $T_f$=20 ms and $T_d$=11.5 ms.

There are four phases in operating the ISCAC. The $^{87}$Rb atoms are first cooled by diffuse light ($T_c$=95 ms), and most of the cold atoms are accumulated in the central area of the microwave cavity [6]. A short pulse ($T_p$=0.5 ms) of pumping light is then injected into the cavity to populate the atoms from F=2 to F=1 state. Then the cold atoms are interrogated by two microwave pulses ($\tau$=3 ms) separated by a free evolution process ($T_f$=20 ms). At last, a probe light ($T_d$=11.5 ms) which is locked to the transition of $5\,^2S_{1/2}\,|F=2\rangle \rightarrow 5\,^2P_{3/2}\,|F'=3\rangle$ is used to detect the population of the state $|F=2, m_F=0\rangle$. The time sequence and servo loop of the clock system are shown in figure 2.

The cycle time of the ISCAC is 133 ms, rather than 92 ms used before [4]. This is because large noises near 10 Hz, which can degrade the clock performance (AM noise), appeared in the spectral density of the Relative Intensity Noise (RIN) of the probe laser (figure 3). However, the atoms' free evolution time Tf keeps the same value of 20 ms. This is a trade-off between narrow line-width and large signal.

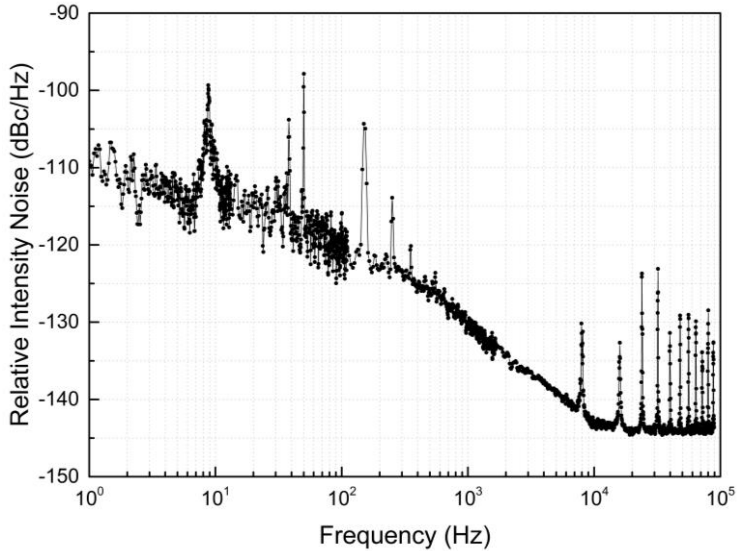

**Figure 3.** The Relative Intensity Noise of the background probe laser after passing through the physical package.

During the Ramsey interrogation process, the two microwave pulses are provided by a low phase noise frequency synthesizer (figure 2). Using the frequency multiplier chain in the frequency synthesizer, the 5 MHz signal coming from the output of the frequency auxiliary generator which is locked to the LO by the inner phase-locked loop is synthesized to 6.834 GHz. Through the clock signal detection at each side of the full width at half maximum (FWHM) of the Ramsey fringe, we can obtain the error signals to steer the output of the frequency auxiliary generator by serial communication. Thus, the noise from the analog feed-back loop which has an impact on the short term frequency stability of about $1.8 \times 10^{-13} \tau^{-1/2}$ can be eliminated.

## 3. RESULTS AND DISCUSSIONS

Figure 4 shows the Allan deviations of the ISCAC. The black curve ($7.3 \times 10^{-13} \tau^{-1/2}$) is the result we have measured [4] before these modifications while the red curve ($5.0 \times 10^{-13} \tau^{-1/2}$) is the frequency stability of the improved clock. The improvement on the short term stability is mainly due to the increasing of the clock signals and the switch of feeding the LO with a digital control loop.

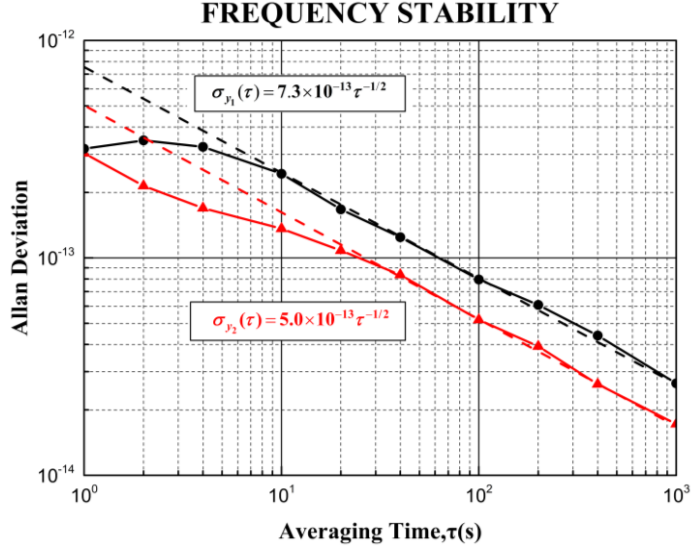

**Figure 4.** The short term frequency stability of the integrating sphere cold atom clock. The red curve which shows better frequency stability is the result after the modifications on the clock.

Limited by the laser power, the previous optical system can only provide total power of about 165 mW as the cooling light which didn't reach the saturation intensity of the system. With the new laser power amplifier module, we can increase the cooling light power to 240 mW. The time of laser cooling stage is also increased from 55 ms to 95 ms. Thus, the intensity of the clock signals changed to about two times of the previous results. Figure 5 shows the intensities of the central Ramsey fringes under these two experimental conditions.

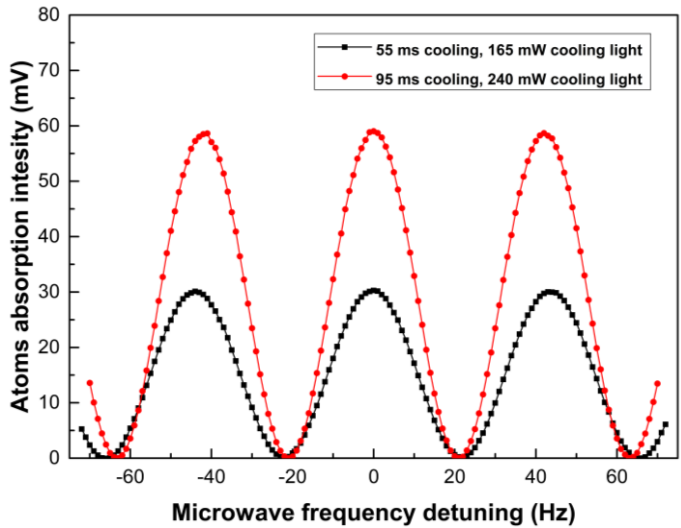

**Figure 5.** The central Ramsey fringes of the clock under different experimental conditions where the curve with red dots is the recent result.

The shot noise of the cold atoms absorption is proportional to $1/\sqrt{N}$ where N is the number of the cold atoms experience the "0-0" transition between the two ground states. Thus, the limitation of the short term frequency stability of atomic shot noise will decrease at the mean time. The shot noise contribution to the frequency stability of the clock is given as [7]

$$\sigma^{sn}_y(\tau) = \frac{1}{\pi Q R_{sn}} \sqrt{\frac{T}{\tau}} \qquad (1)$$

where Q is the quality factor of the atomic resonance, T is the cycle time and $R_{sn}$ is the signal to noise ratio. Table 1 is the list of the noise budget and their instabilities which have impact on the short term frequency stability of the ISCAC.

Table 1. Noise sources of the ISCAC and their instabilities.

| Noise sources | Instability $\sigma_y(\tau)$ |
| --- | --- |
| AM and FM noises of probe laser | $2.0 \times 10^{-13} \tau^{-1/2}$ |
| Atomic shot noise | $3.2 \times 10^{-13} \tau^{-1/2}$ |
| Local oscillator phase noise | $3.3 \times 10^{-13} \tau^{-1/2}$ |
| Total $\sqrt{\sum_i \sigma_{y_i}^2(\tau)}$ | $5.0 \times 10^{-13} \tau^{-1/2}$ |

## 4. CONCLUSIONS

In this paper, we reported the recent progress in the ISCAC. A short term frequency stability of $5.0 \times 10^{-13} \tau^{-1/2}$ has been achieved after we increased the power of the cooling light and used a digital control servo loop to eliminate the noise from the previous analog feedback loop. The intensities of clock signals have increased almost two times which directly lead to a better performance on the short term frequency stability of the cold atom clock.

## REFERENCES

[1] Esnault F X, Holleville D, Rossetto N, Guerandel S, and Dimarcq N, *Phys. Rev. A* **82**, 033436 (2010)
[2] Guillot E, Pottie P E, and Dimarcq N, *Opt. Lett.* 26, 1639 (2001)
[3] Cheng H D, Zhang W Z, Ma H Y, Liu L, and Wang Y Z, *Phys. Rev. A* **79**, 023407 (2009)
[4] Liu P, Meng Y L, Wan J Y, Wang X M, Wang Y N, Xiao L, Cheng H D and Liu L, *Phys. Rev. A* **92**.062101 (2015)
[5] Meng Y L, Zheng B C, Liu P, Wan J Y, Xiao L, Wang X M, Gao Y C, Cheng H D, and Liu L, *Acta Optica Sinica* **34**, 0902001 (2014)
[6] Meng Y L, Cheng H D, Zheng B C, Wang X C, Xiao L, and Liu L, *Chin. Phys. Lett.* **30**, 063701 (2013)
[7] Micalizio S, Godone A, Levi F and Calosso C, *Phys. Rev. A* **79**, 013403 (2009)

## ACKNOWLEDGMENT

This work was supported by National High Technology Research and Development Program of China (No. 2012AA120702).